\documentclass[apj]{emulateapj}

\usepackage{natbib}
\newcommand{\msini}{\ensuremath{M \sin{i}}}
\newcommand{\feh}{\ensuremath{[\rm{Fe}/\rm{H}]}}

\newcommand{\mv}{\ensuremath{M_{\mbox{\scriptsize V}}}}
\newcommand{\dms}{\ensuremath{\Delta \mv}}

\newcommand{\mjup}{\ensuremath{\rm{M}_{\rm{Jup}}}}

\shorttitle{Frequency of Hot Jupiters}
\shortauthors{Wright et al.}

\begin{document}

\title{The Frequency of Hot Jupiters Orbiting Nearby Solar-Type Stars\altaffilmark{\ensuremath{\dagger}}}
\author{J. T. Wright\altaffilmark{1}} \affil{Department of
  Astronomy, 525 Davey Lab, The Pennsylvania State University,
  University Park, PA, 16802\\jtwright@astro.psu.edu} 
\author{G. W. Marcy, A. W. Howard} \affil{Department
  of Astronomy, University of California, Berkeley, CA}
\author{John Asher Johnson\altaffilmark{2}, T. Morton} \affil{Department of Astronomy, California Institute of
  Technology, MC 249-17, Pasadena, CA}
\and
\author{D. A. Fischer} \affil{260 Whitney Avenue, Yale University, New Haven, CT 06511}
\altaffiltext{\ensuremath{\dagger}}{Based on observations obtained
at the W. M. Keck Observatory, which is operated jointly by the
University of California and the California Institute of Technology.
}
\altaffiltext{1}{Center for Exoplanets and Habitable Worlds, The
  Pennsylvania State University, University Park, PA 16801}
\altaffiltext{2}{NASA Exoplanet Science Institute (NExScI), CIT Mail Code 100-22,
770 South Wilson Avenue, Pasadena, CA 91125}

\begin{abstract}
We determine the fraction of F, G, and K dwarfs in the Solar Neighborhood
hosting hot jupiters as measured by the California Planet Survey from
the Lick and Keck planet searches.  We find the rate to be $1.2 \pm
0.38 \%$, which is consistent with the rate reported by Mayor et al.\ (2011) from the
HARPS and CORALIE radial velocity surveys.  These numbers are 
more than double the rate reported by Howard et al.\ (2011) for {\it
  Kepler} stars and the rate of Gould et al.\ (2006) from the OGLE-III
transit search, however due to small number statistics these
differences are of only marginal statistical significance.  We explore
some of the difficulties in  estimating this rate from the existing
radial velocity data sets and comparing radial velocity rates to rates
from other techniques.

\end{abstract}

\section{Introduction}

Hot jupiters are rare objects, a fact obscured by the relative ease
with which they seem to be detected today, and the attention they deservedly
gather.  The first exoplanet discovered orbiting a sun-like star, 51 Peg
{\it b}, was such a close-in giant planet \citep{Mayor_queloz}, and many of the most interesting and 
informative members of the exoplanet menagerie are transiting hot
jupiters.  As a result of the success of ground-based transit surveys, which are
sensitive to almost no other kind of planet, the fraction
of hot jupiters among all known planets is approximately 20\%
\citep[as determined from the Exoplanet Orbit Database,][]{EOD}.  While the overall occurrence rate of planets can be  
high \citep[with 25\% or more of metal-rich or massive stars having
detected planets; ][]{Santos01b,Fischer05c,Johnson10b}, only 
$\sim 7\%$ of radial-velocity-detected planets are hot jupiters.  

The {\it Kepler} mission \citep{Borucki10} has independently measured the rate of
transiting hot jupiters orbiting 58,041 stars, which
\citet{Howard11c} find to be $5 \pm 1$ per thousand stars.  They note
that this is only 40\% of the rate of $1.2 \pm 0.1\%$ (or $12 \pm 1$ per
thousand dwarfs) found by \citet{Marcy_Japan_05} from radial velocity searches. 

The overall rate of hot jupiters is an important constraint on
theories of their origin, as are any differences in the hot jupiter rate
among various stellar populations.  It is therefore interesting and
potentially significant that the rate reported by radial velocity
surveys appears to be in strong conflict with the rate reported by
{\it Kepler} .  

Herein, we perform a new analysis of the hot jupiter occurrence rate
from the Lick and Keck planet searches for comparison with the {\it
  Kepler}, and compare this rate to that found in other radial
velocity and transit searches.

\section{The CPS sample (the denominator) }
\label{denominator}

The California Planet Search (CPS) has been operating since 1988 at Lick
Observatory and since 1995 at Keck Observatory.  The initial target
lists comprised stars carefully selected to be bright, single,
chromospherically quiet, and to span a range of spectral types.  We
have added target lists over the years through many programs,
including programs to monitor low-mass stars \citep[the
Keck M-dwarf survey and M2K; ][]{Johnson10,Apps10}, intermediate mass stars \citep[``Retired A
Stars'';][]{Johnson07}, high metallicity stars \citep[N2K;][]{Fischer05}, active stars, SIM reference stars, Kepler
targets, systems discovered by transit \citep[e.g.\ the HATNet targets;][]{Bakos07}, planetary systems discovered by
radial velocities (RVs) by other groups, and many other purposes.   Today,
this master list of targets has been observed at a variety of
cadences, resulting in a range of sensitivities to exoplanets that
varies strongly with planetary parameters and from star to star.

Fortunately, the detection of hot jupiters with precise radial
velocities does not require intense observation.  \citet{Fischer05} showed
that hot jupiters could be reliably identified with only 3 or 4 observations
spaced over a few days, since their periods are short and amplitudes
are high.  We can thus crudely but not inaccurately identify
those stars for which we could detect hot jupiters, should they
exist, simply from the number of observations we have made.  This metric is not perfect;  stars discovered to be
spectroscopic binaries or highly active (and thus having large
stochastic RV variations) might be observed only 5 or 6
times before being dropped from our program and yet still harbor an
undetected hot Jupiter.  Based on our extensive familiarity with our
program, we judge the number of such systems in the
following discussion to be an insignificant contributor to our error
budget.  

Since most stars in our sample have been observed many more
than 6 times, we are essentially complete to hot jupiters.  In principle,
we will have some very small contamination from face-on binaries and
miss some number of face-on hot jupiters.  These effects work in
opposite directions and are both very small compared to the Poisson
noise in our sample;  we neglect them here.

We cannot construct a target list that is perfectly statistically 
matched to the {\it Kepler} targets, not least because those targets
have not been perfectly characterized; the Kepler Input Catalog
provided crude temperatures and luminosity classes but is not, and was
never intended to be, a precise tool \citep{KIC}.  Nonetheless, we can make a somewhat
clean comparison to the {\it Kepler} sample by mimicking certain aspects of
that survey.  

We have first performed a magnitude cut at $V < 8$ (we use
magnitudes from Hipparcos \citep{PerrymanESA} throughout this work).
This cut removes all stars added to our target list as part of
follow-up to transit search programs (the brightest 
transit-discovered planet, WASP-33, has $V=8.3$ \citep{Christian06}).
This magnitude cut is near our completeness limit for predominantly
old, cool, and single dwarf and subgiant stars easily accessible from Lick and
Keck Observatories.  This also generates a Malmquist bias
\citep{Malmquist} that favors evolved and metal-rich stars at 
any given color.  We discuss the degree to which this bias also exists
in the other samples we consider in \S\ref{compare}.

In order to make a fair comparison with other works, we perform a
color cut at $B-V < 1.2$ to exclude very cool stars, which are not
well represented in the {\it Kepler}  or CORALIE samples  (though the
\citet{Gould06} result does include these lower-mass stars).  We note
that the giant planet frequency around M dwarfs appears to be lower than that
of FGK stars \citep{Johnson10b,Bonfils11}, and so it is appropriate to exclude
them here. 
   
Finally, we perform an evolution cut, including only dwarf and subgiant 
stars (those whose height above the Main Sequence $\dms < 2.5$, using
the Main Sequence fit of \citet{WrightErratum05}).

Of the remaining stars on our target list, 836 have a number of
observations at any one telescope $N_{\rm{obs}} \geq 5$.  Four
observations should be sufficient, in principle, to detect a hot
jupiter;  we explore the sensitivity of our results to
$N_{\rm{obs}}$ below.

\section{The Hot Jupiters (the numerator) }
\label{numerator}

We use the Exoplanet Orbit Database at exoplanets.org \citep{EOD} to
determine the number of planets in our sample.  The EOD contains only
planets with well determined orbits described in peer-reviewed
journals. We define a planet as being a hot jupiter in our sample if
it orbits one of the stars in our denominator (\S~\ref{denominator}),
the EOD lists it, and it has $P < 10$ d and $\msini > 0.1 \mjup$.    We
list all planets meeting these criteria in Table 1.

Because {\it Kepler} measures planetary radius, and not mass, a
perfect comparison of the samples is 
not possible until all of the {\it Kepler} planets have masses
measured through radial velocity or other studies.   We have chosen 0.1 \mjup\
as a lower mass limit for ``hot jupiter'' to 
most closely match the \citet{Cumming08} analysis, and because
the term ``hot jupiter'' has traditionally referred to such massive
planets, and not to presumed ``ice giants'' with $M < 0.1 \mjup$.

This mass limit corresponds to a radius limit of 8 $R_\Earth$ as
adopted by \citet{Howard11c} for densities of 1.4 g/cc, typical
of gas giants with small rocky cores.  Clearly, it is impossible to
associate a given mass limit with a precise radius limit, as the
densities of such planets range from 0.1 -- 1.4 g/cc.  But lower
densities imply larger radii, and so all such planets will included in the
\citet{Howard11c} sample of hot jupiters from {\it Kepler}.  

To get a sense of how our sample limits affect our statistics, we have
divided hot jupiter hosts into two samples.  Stars in our primary sample
satisfy our stellar cuts and were 
in our Keck and Lick samples before the discovery of planets orbiting
them.  They unambiguously represent hot jupiters detected in our
sample.

Two special cases are 51 Peg and HD 189733.  While 51 Peg was not on
our original Lick target list, the star is 
sufficiently bright that it would certainly have eventually been included in the
Lick and Keck planet searches, even if its planet had not been discovered
by \citeauthor{Mayor_queloz}.   Similarly, while
HD 189733 was not first announced by our team, we were following the 
target and had detected the planet when \citet{Bouchy05b} announced
it.  It would be thus inappropriate to ignore these planets from our
statistics, and so we include them in our primary sample.

Stars in our expanded sample are those just beyond our cutoff, extending into
$8.0 < V \leq 8.1$ and $1.2 < B-V < 1.25$.  The number of stars in our
sample that we would have under this expanded definition is $N_{\rm{obs}} \geq \{4,5,6\} =
\{965,906,843\}$.   There are 5 extended sample hot jupiters in Table 1.

For completeness, we also include in Table 1 the hot Jupiters orbiting stars
that lie outside our sample's various cuts, or that we excluded
because their hosts were added to our target list only after a planet was
discovered by another team.  

\begin{deluxetable*}{llllllll}
\tabletypesize{\scriptsize}
\tablecolumns{8}
\tablecaption{Radial Velocity Detected Hot Jupiters}
\tablehead{\colhead{Planet} & \colhead{\msini(\mjup)} & \colhead{B-V}
  & \colhead{V} & \colhead{$P$ (d)} &
  \colhead{distance (pc)} & \colhead{First Reference\tablenotemark{1}} & \colhead{Sample\tablenotemark{2}}}
\startdata
$\upsilon$ And {\it b}& 0.669 $\pm$ 0.026 & 0.536 & 4.1 & 4.6 & 13.492 $\pm$ 0.035 & \citet{Butler97} &  in primary sample\\
$\tau$ Boo {\it b}& 4.12 $\pm$ 0.15 & 0.508 & 4.5 & 3.3 & 15.618 $\pm$ 0.046 & \citet{Butler97} &  in primary sample\\
51 Peg {\it b}& 0.461 $\pm$ 0.016 & 0.666 & 5.45 & 4.2 & 15.608 $\pm$ 0.093 & \citet{Mayor_queloz} &  included in primary sample\\
HD 217107 {\it b}& 1.401 $\pm$ 0.048 & 0.744 & 6.17 & 7.1 & 19.86 $\pm$ 0.15 & \citet{Fischer99} &  in primary sample\\
HD 185269 {\it b}& 0.954 $\pm$ 0.069 & 0.606 & 6.67 & 6.8 & 50.3 $\pm$ 
1.4 & \citet{Johnson06b} &  in primary sample\\ 
 HD 209458 {\it b}& 0.689 $\pm$ 0.024 & 0.594 & 7.65 & 3.5 & 49.6 $\pm$ 2.0 & \citet{HenryG00,Charbonneau00} &  in primary sample\\ 
HD 189733 {\it b}& 1.140 $\pm$ 0.056 &0.931 & 7.67 & 2.2 & 19.45 $\pm$ 0.26
& \citet{Bouchy05b} &  included in primary sample\\  
HD 187123 {\it b}& 0.510 $\pm$ 0.017 & 0.661 & 7.83 & 3.1 & 48.3 $\pm$ 1.2 & \citet{Butler98} &  in primary sample\\ 
HD 46375 {\it b}& 0.2272 $\pm$ 0.0091 & 0.860 & 7.91 & 3.0 & 34.8 $\pm$ 1.1 & \citet{Marcy_saturns} &  in primary sample\\
HD 149143 {\it b}& 1.328 $\pm$ 0.078 & 0.714 & 7.89 & 4.0 & 62.0 $\pm$ 3.2 &
\citet{Fischer06,DaSilva06} &  in primary sample \\

\hline

HD 88133 {\it b}& 0.299 $\pm$ 0.027 & 0.810 & 8.01 & 3.4 & 81.4 $\pm$ 5.8 &
\citet{Fischer05} &  in expanded sample \\
HD 102956 {\it b}& 0.955 $\pm$ 0.048 & 0.971 & 8.02 & 6.5 & 126 $\pm$ 13 &
\citet{Johnson10c} &  in expanded sample \\
HD 109749 {\it b}& 0.275 $\pm$ 0.016 & 0.680 & 8.08 & 5.2 & 56.3 $\pm$ 4.0 & \citet{Fischer06} &  in expanded sample\\ 
HD 49674 {\it b}& 0.1016 $\pm$ 0.0082 & 0.729 & 8.1 & 4.9 & 44.2 $\pm$ 1.7 &
\citet{Butler03} &  in expanded sample \\

\hline

HD 179949 {\it b}& 0.902 $\pm$ 0.033 & 0.548 & 6.25 & 3.1 & 27.55 $\pm$ 0.53 & \citet{Tinney01} & excluded from primary sample \\
HD 168746 {\it b}& 0.245 $\pm$ 0.017 & 0.713 & 7.95 & 6.4 & 42.7 $\pm$ 1.4 & \citet{Pepe02} & excluded from primary sample\\
HD 102195 {\it b}& 0.453 $\pm$ 0.021 & 0.835 & 8.07 & 4.1 & 29.64 $\pm$ 0.73 & \citet{Ge06} & excluded from expanded sample\\ 
HD 73256 {\it b}& 1.869 $\pm$ 0.083 & 0.782 & 8.08 & 5.2 & 37.76 $\pm$ 0.91 & \citet{Udry03} & excluded from expanded sample\\
HD 149026 {\it b}& 0.360 $\pm$ 0.016 & 0.611 & 8.16 & 2.9 & 79.4 $\pm$ 4.4 & \citet{Sato05} & outside sample \\ 
HD 68988 {\it b}& 1.80 $\pm$ 0.10 & 0.652 & 8.2 & 6.3 & 54.5 $\pm$ 2.3 & \citet{Vogt02} & outside sample \\
HD 83443 {\it b}& 0.396 $\pm$ 0.018 & 0.811 & 8.23 & 3.0 & 41.2 $\pm$ 1.2 & \citet{Butler02} & outside sample \\
HIP 14810 {\it b}& 3.87 $\pm$ 0.13 & 0.777 & 8.52 & 6.7 & 53.4 $\pm$ 3.6 & \citet{Butler06,Wright09c} & outside sample \\
HD 86081 {\it b}& 1.496 $\pm$ 0.050 & 0.664 & 8.73 & 2.1 & 95.3 $\pm$ 9.0 & \citet{Johnson06a} & outside sample \\
BD -10 3166 {\it b}& 0.430 $\pm$ 0.017 & 0.903 & 10.08 & 3.5 & 80.0 $\pm$ 8.0 & \citet{Butler00} & outside sample
\enddata
\tablenotetext{1}{First refereed source of orbital elements; taken from the
  Exoplanet Orbit Database \citep{EOD}.}
\tablenotetext{2}{See text for more specific sample definitions}

\end{deluxetable*}

\section{The Frequency of Hot Jupiters (the ratio)}

We choose to use only primary sample stars with 5 or more measurements for
our analysis, yielding a hot jupiter frequency of 10/836 or $12.0 \pm
3.8$ per thousand stars\footnote{Our analysis is simplified by the
  fact that there are no systems in our sample with {\it multiple} hot jupiters.}. 

Our sample cutoffs in color, evolutionary state, and magnitude were chosen to be
round numbers.  The inclusion of the 906 stars with $N_{\rm{obs}} \geq
5$ from the extended sample (and their planets) 
gives a sense of how sensitive our numbers are to these limits.    Our
expanded sample then includes 4 additional planets, and the implied 
rate in our expanded sample is thus $15.5 \pm 4.1$ per thousand.  We
note that HD 109749 and HD 88133 were added to our sample as part of the 
N2K program, which targeted metal-rich (and thus planet-rich) stars, and
HD 102956 was added as part of the ``Retired A Stars'' program which,
in retrospect, similarly targets planet-rich stars, though they may
lack many hot jupiters.  This expanded sample, which shows an
additional 3.5 hot jupiters per thousand stars, is thus probably not
representative of field FGK dwarfs, which explains its slightly
higher hot jupiter detection fraction.  

Finally, we can vary the minimum number of observations we require of
a star in order for our RV survey to be sensitive to a hot jupiters.  The number of stars in
our sample is $\{890,836,785\}$ for $N_{\rm{obs}} \geq \{4,5,6\}$, implying a hot
jupiter rate of $\{11.2,12.0,12.7\}$ for these values.  From the spread
in these values we estimate a ``systematic error'' of $\sim 0.7$ per thousand stars from
this consideration.  

We thus estimate that the true rate of hot
jupiter detections around FGK dwarfs and subgiants in our sample is
$1.20 \pm 0.38 \%$, with some small, additional contribution of ``systematic'' 
error of order $0.07\%$ from our choices for $N_{\rm{obs}}$,
and some potentially larger systematic error stemming from our sample
cuts.  Certainly this rate could be more robustly determined, but we note that the
random Poisson errors here are at least as large as these systematic errors, and so
the latter probably do not warrant significant further refinement.  

\section{Comparison With Earlier Results and Other Surveys}
\label{compare}

We summarize the hot jupiter rates implied by various surveys and
published analyses in Table 2, and describe them in more detail below

\subsection{Transit Surveys}

Transit surveys are not, of course, complete to hot jupiters because
they require edge-on geometry, but the assumption of isotropy of the
ensemble of
orbital planes makes the calculation a true hot jupiter rate
straightforward \citep[but hardly trivial, see, e.g.][]{Gaudi05a}.  The
most thorough calculations of the transit-survey hot-jupiter rate are those
from OGLE-III and {\it Kepler}.

Both surveys may probe a significantly different population than the
radial velocity surveys.  For instance, based on stellar population models of the Milky Way, \citet{Gould06} calculate that the magnitude limits imposed in transit surveys may produce a sample with a significantly different metallicity distribution than would be seen in an RV survey of nearby stars.  They estimate that, compared to the true metallicity distribution in the Galaxy, RV survey samples will be overrepresented by 20\% for every 0.1 dex in \feh\ from the Malmquist bias if magnitude cuts
are made in discrete bins of $B-V$.  They predict that the OGLE-III
sample should exhibit a similar but smaller overrepresentation of
metal-rich stars of 2\% per dex in \feh.  If this is correct, then transit surveys like OGLE and {\it Kepler} probe a lower-metallicity population, on average, than radial velocity surveys.

\subsubsection{OGLE-III}

\citet{Gould06} reported the hot jupiter rate implied by the OGLE-III
transit survey to be $3.1^{+4.3}_{-1.8}$ (90\% confidence limits) hot jupiters per
thosand stars.  In this study a ``hot jupiter'' was any detection with $P
= 3$--5 d.  OGLE surveyed 52,000 stars toward the Galactic Center and
103,000 stars toward Carina, with sensitivity to planets down to
$\sim 1 R_{\rm Jup}$ , and rapidly decreasing sensitivity below this level.
Like in our analysis, \citet{Gould06} restricted their calculation to
main sequence stars, but with a cutoff of $V < 17.5$ mag (while many subgiants and
giants were observed in the survey, the larger radii of these stars
make planet detection around them impossible and so they were
not considered).  If we calculate our hot Jupiter rate using a similar
cutoff of $P < 5$ d, we find a rate of 9.6 per thousand, still over 3
times the OGLE rate.

\subsubsection{{\it Kepler}}

\citet{Howard11c} reported that the occurrence rate from {\it Kepler}
based on an anlysis of 58,041 GK dwarfs in the larger 156,000 {\it
  Kepler} sample.  They find the frequency of giant ($R_P = 
$8--32$ R_\Earth$) planets with periods $P < 10$ d to be $4\pm 1$ per
thousand stars of magnitude $K_p < 15$, and $5 \pm 1$ per
thousand stars with $K_p < 16$.   

The {\it Kepler} field is centered at $b = +13.3^\circ$, so the
distant stars it probes (sitting several hundreds of parsecs from
Earth) will have significant heights above the galactic plane,
potentially distinguishing that
population from that in the radial velocity surveys by age and,
possibly, metallicity.

\subsection{Radial Velocity Surveys}

The radial velocity surveys described here can be roughly divided into
two broad collaborations, each encompassing multiple programs and
teams: efforts by members of the Keck, Lick, and Anglo-Australian
Planet Searches, and  the European / Geneva efforts with the ELODIE,
CORALIE, SOPHIE, and HARPS spectrographs.  The target lists of these
searches have some overlap (since they target the brightest stars) but
their methodologies and analysis procedures are independent.  We
describe the most important three prior measurements of the hot
jupiter rate around nearby dwarfs here.

\subsubsection{Marcy et al.\ (2005)}

The overall rate of hot jupiters among FGK dwarfs surveyed by radial
velocity was estimated by \citet{Marcy_Japan_05} to be $1.2 \pm 0.1\%$$(12 \pm 1$ per
thousand).  This study analyzed 1330 stars from the
Lick, Keck, and Anglo-Australian Planet Searches, and counted the
number of detections of planets of any minimum mass with semimajor
axis $a < 0.1$ AU.  The \citeauthor{Marcy_Japan_05} study is thus
similar to our analysis here in that it employs a similar sample of
stars and radial velocities, but has some methodological differences. 

\citeauthor{Marcy_Japan_05}\ used a well defined initial sample, but
the planets included had no specified cutoff at $\msini > 0.1 \mjup$,
and so included a small number of planets with significantly lower
masses than those considered here \citep[and presumably significantly
smaller than the 8 $R_\Earth$ cutoff used by][]{Howard11c}.
\citeauthor{Marcy_Japan_05} also included the AAT planet search, where
we do not.  Despite those differences, we find an identical result in
our analysis, but with more conservative uncertainties.

\subsubsection{Cumming et al.\ (2008)}

\citet{Cumming08} performed a careful analysis of the detectability
of target stars in the Keck planets search to determine the
distribution of planets in the minimum-mass-period plane, and found
that $15\pm 6$ stars in a thousand harbor a planet with $\msini > 0.3
\mjup$ and $P < 11.5$ d, and $20\pm 7$ stars in a thousand for planets
with $\msini > 0.1\mjup$. 

\citeauthor{Cumming08}, however, made no attempt to distinguish amongst stars added
blindly and stars added because they were more likely to have planets
(metal rich stars), or specifically because they had some property not
representative of the {\it Kepler} field (i.e.\ subgiants), or even
because they were already known to have planets.  In other words,
\citeauthor{Cumming08} determined the hot jupiter frequency in the
Keck sample, but that sample is clearly enriched with respect to the
field, which explains their higher hot jupiter rate.  We have avoided
these effects with out various sample cuts (see \S\S~\ref{denominator},\ref{numerator}).

\subsubsection{Mayor et al.\ (2011)}

Most recently, \citet{Mayor11} used the HARPS and CORALIE RV planet survey to estimate 
the hot jupiter occurrence rate as a function of minimum mass and
period in their sample of dwarf stars.  Their occurrence rate for planets with $\msini > 50 M_\Earth$
and $P < 11$ d is $8.9 \pm 3.6$ per thousand stars, which is consistent
with both {\it Kepler} and the \citeauthor{Marcy_Japan_05}
results.  The different minimum $\msini$ and maximum period
used between the studies of \citeauthor{Marcy_Japan_05} and \citeauthor{Mayor11} not
significant here because there is only one planet in the 
  Exoplanet Orbit Database with $30 M_\Earth < \msini < 50 M_\Earth$
  and $P < 10$ d (HD 49674$b$) and only one with both $10 < P < 11$ d
  and $V<8$, and neither host star is in our primary sample.  Our statistics
  would thus be identical if we had adopted the cutoffs of
  \citeauthor{Mayor11} instead of emulating those of \citeauthor{Howard11c}.

\section{Discussion and Conclusion}

The analysis presented here fairly represents, within the Poisson
noise, the true hot jupiter frequency among old FGK dwarfs in the
Solar Neighborhood.  We note that our result is not enhanced by high
metallicity stars from the N2K sample, since most of those stars are
fainter than those in our primary sample, and especially since only one of our
primary sample planets was discovered by that survey (HD 149143).    

We decline here to attempt a more rigorous comparison the Keck and
Lick stellar sample, with its heterogeneous selection effects and
sensitivities, to the {\it Kepler} sample.  {\it Kepler}, as noted
above, doubtless probes a different 
stellar population than ours, since those stars were selected from
different criteria, lie at a different Galactocentric radius and
Galactic height, and so have a different distribution of ages,
evolutionary states, binary fractions, and dynamical histories.  We
also note that our binary star rejection 
predominently rejects binaries with separations $< 2^{\prime\prime}$
due to concerns of spectral contamination at the slit.  Transit
surveys will typically have no such rejection criterion, though they
may have a more difficult validation or confirmation procedure for
binary stars.

Nonetheless, it is interesting to note that there is some indication
that the RV surveys, which probe the Solar Neighborhood, are
consistently finding a hot jupiter rate at least twice that seen with the transit
surveys.  Given the differences in the
stellar populations of these surveys, this is perhaps not surprising.  
 
The most salient difference between the samples may be metallicity.
If the difference in metallicity bias between the RV and transit samples is as severe as
\citet{Gould06} estimate, then the difference in the hot jupiter rate may
simply reflect the difference in giant planet occurrence rate among high and
low metallicity stars.  We point out that the \citet{Gould06} estimate
is based on stellar population models of the Milky Way, and not on a
comparison of metallicity measurements between transit and RV samples.
The stars in the RV samples are generally well known and well
characterized, because the stars are known to be single, generally
have good parallaxes, and have metallicities measured from spectra and
color-absolute magnitude information.  A thorough spectroscopic
metallicity analysis of a statistically appropriate sample of {\it
  Kepler} targets should provide a similar sense of the {\it Kepler}
metallicity distribution, which would help confirm or rule out
metallicity as a source of the hot jupiter rate discrepancy. 

We close noting that the apparent RV vs.\ transit hot jupiter rate
discrepancy, while apparently large, is of only marginal 
statistical significance:  the hot jupiter frequency per thousand
stars of the Keck and Lick sample (12.0 $\pm 3.8$) and from
the \citeauthor{Mayor11} sample  ($8.9 \pm 3.6$) is only 1--2-$\sigma$
discrepant from  the \citeauthor{Gould06} frequency from OGLE-III
transits ($3.1^{+4.3}_{-1.8}$, 90\% confidence limits) and with the frequency of $5 \pm 1$ in the {\it
  Kepler} sample found by \citeauthor{Howard11c}

\begin{deluxetable*}{lll}
\tabletypesize{\scriptsize}
\tablecolumns{3}
\tablecaption{Hot Jupiter Rate from Previous Works}
\tablehead{\colhead{Work} & \colhead{Rate (per thousand)} & \colhead{Sample}}
\startdata
\citet{Gould06} & $3.1^{+4.3}_{-1.8}$ & OGLE-III Transits (90\%
confidence limits, $P < 5$ d) \\
\citet{Howard11c} &  $5 \pm 1$& Kepler Transits\\
\hline
\citet{Marcy05} & $12 \pm 1$ & Keck, Lick, and AAT RVs \\
\citet{Cumming08} & $15 \pm 6$ & Keck RVs (Entire target list)\\
\citet{Mayor11} &  $8.9 \pm 3.6$ & HARPS and CORALIE RVs \\
This Work &  $12.0 \pm 3.8$ & Keck and Lick RVs
\enddata
\end{deluxetable*}

\acknowledgments
This work was partially supported by funding from the Center for Exoplanets
and Habitable Worlds, which is supported by the Pennsylvania State University, the Eberly College
of Science, and the Pennsylvania Space Grant Consortium.  

The work herein is based on observations obtained at the W. M. Keck
Observatory, which is operated jointly by the University of California
and the California Institute of Technology.  The Keck Observatory was
made possible by the generous financial support of the W.M. Keck
Foundation.  We wish to recognize and acknowledge the very significant
cultural role and reverence that the summit of Mauna Kea has always
had within the indigenous Hawaiian community.  We are most fortunate
to have the opportunity to conduct observations from this mountain.

This research has made use of the Exoplanet Orbit Database
and the Exoplanet Data Explorer at exoplanets.org.

\facility{{\it Facility} Keck:I}


\begin{thebibliography}{47}
\expandafter\ifx\csname natexlab\endcsname\relax\def\natexlab#1{#1}\fi

\bibitem[{{Apps} {et~al.}(2010){Apps}, {Clubb}, {Fischer}, {Gaidos}, {Howard},
  {Johnson}, {Marcy}, {Isaacson}, {Giguere}, {Valenti}, {Rodriguez}, {Chubak},
  \& {Lepine}}]{Apps10}
{Apps}, K., {et~al.} 2010, \pasp, 122, 156

\bibitem[{{Bakos} {et~al.}(2007){Bakos}, {Noyes}, {Kov{\'a}cs}, {Latham},
  {Sasselov}, {Torres}, {Fischer}, {Stefanik}, {Sato}, {Johnson}, {P{\'a}l},
  {Marcy}, {Butler}, {Esquerdo}, {Stanek}, {L{\'a}z{\'a}r}, {Papp}, {S{\'a}ri},
  \& {Sip{\H o}cz}}]{Bakos07}
{Bakos}, G.~{\'A}., {et~al.} 2007, \apj, 656, 552

\bibitem[{{Bonfils} {et~al.}(2011){Bonfils}, {Delfosse}, {Udry}, {Forveille},
  {Mayor}, {Perrier}, {Bouchy}, {Gillon}, {Lovis}, {Pepe}, {Queloz}, {Santos},
  {S{\'e}gransan}, \& {Bertaux}}]{Bonfils11}
{Bonfils}, X., {et~al.} 2011, ArXiv e-prints

\bibitem[{{Borucki} {et~al.}(2010){Borucki}, {Koch}, {Basri}, {Batalha},
  {Brown}, {Caldwell}, {Caldwell}, {Christensen-Dalsgaard}, {Cochran},
  {DeVore}, {Dunham}, {Dupree}, {Gautier}, {Geary}, {Gilliland}, {Gould},
  {Howell}, {Jenkins}, {Kjeldsen}, {Kondo}, {Latham}, {Lissauer}, {Marcy},
  {Meibom}, {Monet}, {Morrison}, {Sasselov}, \& {Tarter}}]{Borucki10}
{Borucki}, W.~J., {et~al.} 2010, in Bulletin of the American Astronomical
  Society, Vol.~41, Bulletin of the American Astronomical Society, 215--+

\bibitem[{{Bouchy} {et~al.}(2005){Bouchy}, {Udry}, {Mayor}, {Moutou}, {Pont},
  {Iribarne}, {da Silva}, {Ilovaisky}, {Queloz}, {Santos}, {S{\'e}gransan}, \&
  {Zucker}}]{Bouchy05b}
{Bouchy}, F., {et~al.} 2005, \aap, 444, L15

\bibitem[{{Brown} {et~al.}(2011){Brown}, {Latham}, {Everett}, \&
  {Esquerdo}}]{KIC}
{Brown}, T.~M., {Latham}, D.~W., {Everett}, M.~E., \& {Esquerdo}, G.~A. 2011,
  \aj, 142, 112

\bibitem[{{Butler} {et~al.}(1998){Butler}, {Marcy}, {Vogt}, \&
  {Apps}}]{Butler98}
{Butler}, R.~P., {Marcy}, G.~W., {Vogt}, S.~S., \& {Apps}, K. 1998, \pasp, 110,
  1389

\bibitem[{{Butler} {et~al.}(2003){Butler}, {Marcy}, {Vogt}, {Fischer}, {Henry},
  {Laughlin}, \& {Wright}}]{Butler03}
{Butler}, R.~P., {Marcy}, G.~W., {Vogt}, S.~S., {Fischer}, D.~A., {Henry},
  G.~W., {Laughlin}, G., \& {Wright}, J.~T. 2003, \apj, 582, 455

\bibitem[{{Butler} {et~al.}(1997){Butler}, {Marcy}, {Williams}, {Hauser}, \&
  {Shirts}}]{Butler97}
{Butler}, R.~P., {Marcy}, G.~W., {Williams}, E., {Hauser}, H., \& {Shirts}, P.
  1997, \apjl, 474, L115+

\bibitem[{{Butler} {et~al.}(2000){Butler}, {Vogt}, {Marcy}, {Fischer}, {Henry},
  \& {Apps}}]{Butler00}
{Butler}, R.~P., {Vogt}, S.~S., {Marcy}, G.~W., {Fischer}, D.~A., {Henry},
  G.~W., \& {Apps}, K. 2000, \apj, 545, 504

\bibitem[{{Butler} {et~al.}(2002){Butler}, {Marcy}, {Vogt}, {Tinney}, {Jones},
  {McCarthy}, {Penny}, {Apps}, \& {Carter}}]{Butler02}
{Butler}, R.~P., {et~al.} 2002, \apj, 578, 565

\bibitem[{{Butler} {et~al.}(2006){Butler}, {Wright}, {Marcy}, {Fischer},
  {Vogt}, {Tinney}, {Jones}, {Carter}, {Johnson}, {McCarthy}, \&
  {Penny}}]{Butler06}
---. 2006, \apj, 646, 505

\bibitem[{{Charbonneau} {et~al.}(2000){Charbonneau}, {Brown}, {Latham}, \&
  {Mayor}}]{Charbonneau00}
{Charbonneau}, D., {Brown}, T.~M., {Latham}, D.~W., \& {Mayor}, M. 2000, \apjl,
  529, L45

\bibitem[{{Christian} {et~al.}(2006){Christian}, {Pollacco}, {Skillen},
  {Street}, {Keenan}, {Clarkson}, {Collier Cameron}, {Kane}, {Lister}, {West},
  {Enoch}, {Evans}, {Fitzsimmons}, {Haswell}, {Hellier}, {Hodgkin}, {Horne},
  {Irwin}, {Norton}, {Osborne}, {Ryans}, {Wheatley}, \& {Wilson}}]{Christian06}
{Christian}, D.~J., {et~al.} 2006, \mnras, 372, 1117

\bibitem[{{Cumming} {et~al.}(2008){Cumming}, {Butler}, {Marcy}, {Vogt},
  {Wright}, \& {Fischer}}]{Cumming08}
{Cumming}, A., {Butler}, R.~P., {Marcy}, G.~W., {Vogt}, S.~S., {Wright}, J.~T.,
  \& {Fischer}, D.~A. 2008, \pasp, 120, 531

\bibitem[{{da Silva} {et~al.}(2006){da Silva}, {Udry}, {Bouchy}, {Mayor},
  {Moutou}, {Pont}, {Queloz}, {Santos}, {S{\'e}gransan}, \&
  {Zucker}}]{DaSilva06}
{da Silva}, R., {et~al.} 2006, \aap, 446, 717

\bibitem[{{Fischer} {et~al.}(1999){Fischer}, {Marcy}, {Butler}, {Vogt}, \&
  {Apps}}]{Fischer99}
{Fischer}, D.~A., {Marcy}, G.~W., {Butler}, R.~P., {Vogt}, S.~S., \& {Apps}, K.
  1999, \pasp, 111, 50

\bibitem[{{Fischer} \& {Valenti}(2005)}]{Fischer05c}
{Fischer}, D.~A., \& {Valenti}, J. 2005, \apj, 622, 1102

\bibitem[{{Fischer} {et~al.}(2005){Fischer}, {Laughlin}, {Butler}, {Marcy},
  {Johnson}, {Henry}, {Valenti}, {Vogt}, {Ammons}, {Robinson}, {Spear},
  {Strader}, {Driscoll}, {Fuller}, {Johnson}, {Manrao}, {McCarthy},
  {Mu{\~n}oz}, {Tah}, {Wright}, {Ida}, {Sato}, {Toyota}, \&
  {Minniti}}]{Fischer05}
{Fischer}, D.~A., {et~al.} 2005, \apj, 620, 481

\bibitem[{{Fischer} {et~al.}(2006){Fischer}, {Laughlin}, {Marcy}, {Butler},
  {Vogt}, {Johnson}, {Henry}, {McCarthy}, {Ammons}, {Robinson}, {Strader},
  {Valenti}, {McCullough}, {Charbonneau}, {Haislip}, {Knutson}, {Reichart},
  {McGee}, {Monard}, {Wright}, {Ida}, {Sato}, \& {Minniti}}]{Fischer06}
---. 2006, \apj, 637, 1094

\bibitem[{{Gaudi} {et~al.}(2005){Gaudi}, {Seager}, \&
  {Mallen-Ornelas}}]{Gaudi05a}
{Gaudi}, B.~S., {Seager}, S., \& {Mallen-Ornelas}, G. 2005, \apj, 623, 472

\bibitem[{{Ge} {et~al.}(2006){Ge}, {van Eyken}, {Mahadevan}, {DeWitt}, {Kane},
  {Cohen}, {Vanden Heuvel}, {Fleming}, {Guo}, {Henry}, {Schneider}, {Ramsey},
  {Wittenmyer}, {Endl}, {Cochran}, {Ford}, {Mart{\'{\i}}n}, {Israelian},
  {Valenti}, \& {Montes}}]{Ge06}
{Ge}, J., {et~al.} 2006, \apj, 648, 683

\bibitem[{{Gould} {et~al.}(2006){Gould}, {Dorsher}, {Gaudi}, \&
  {Udalski}}]{Gould06}
{Gould}, A., {Dorsher}, S., {Gaudi}, B.~S., \& {Udalski}, A. 2006, Acta
  Astronomica, 56, 1

\bibitem[{{Henry} {et~al.}(2000){Henry}, {Marcy}, {Butler}, \&
  {Vogt}}]{HenryG00}
{Henry}, G.~W., {Marcy}, G.~W., {Butler}, R.~P., \& {Vogt}, S.~S. 2000, \apjl,
  529, L41

\bibitem[{{Howard} {et~al.}(2011){Howard}, {Marcy}, {Bryson}, {Jenkins},
  {Rowe}, {Batalha}, {Borucki}, {Koch}, {Dunham}, {Gautier}, {Van Cleve},
  {Cochran}, {Latham}, {Lissauer}, {Torres}, {Brown}, {Gilliland}, {Buchhave},
  {Caldwell}, {Christensen-Dalsgaard}, {Ciardi}, {Fressin}, {Haas}, {Howell},
  {Kjeldsen}, {Seager}, {Rogers}, {Sasselov}, {Steffen}, {Basri},
  {Charbonneau}, {Christiansen}, {Clarke}, {Dupree}, {Fabrycky}, {Fischer},
  {Ford}, {Fortney}, {Tarter}, {Girouard}, {Holman}, {Johnson}, {Klaus},
  {Machalek}, {Moorhead}, {Morehead}, {Ragozzine}, {Tenenbaum}, {Twicken},
  {Quinn}, {Isaacson}, {Shporer}, {Lucas}, {Walkowicz}, {Welsh}, {Boss},
  {Devore}, {Gould}, {Smith}, {Morris}, {Prsa}, \& {Morton}}]{Howard11c}
{Howard}, A.~W., {et~al.} 2011, arXiv:1103.2541 (ApJ, submitted)

\bibitem[{{Johnson} {et~al.}(2010{\natexlab{a}}){Johnson}, {Aller}, {Howard},
  \& {Crepp}}]{Johnson10b}
{Johnson}, J.~A., {Aller}, K.~M., {Howard}, A.~W., \& {Crepp}, J.~R.
  2010{\natexlab{a}}, \pasp, 122, 905

\bibitem[{{Johnson} {et~al.}(2006{\natexlab{a}}){Johnson}, {Marcy}, {Fischer},
  {Henry}, {Wright}, {Isaacson}, \& {McCarthy}}]{Johnson06b}
{Johnson}, J.~A., {Marcy}, G.~W., {Fischer}, D.~A., {Henry}, G.~W., {Wright},
  J.~T., {Isaacson}, H., \& {McCarthy}, C. 2006{\natexlab{a}}, \apj, 652, 1724

\bibitem[{{Johnson} {et~al.}(2006{\natexlab{b}}){Johnson}, {Marcy}, {Fischer},
  {Laughlin}, {Butler}, {Henry}, {Valenti}, {Ford}, {Vogt}, \&
  {Wright}}]{Johnson06a}
{Johnson}, J.~A., {et~al.} 2006{\natexlab{b}}, \apj, 647, 600

\bibitem[{{Johnson} {et~al.}(2007){Johnson}, {Fischer}, {Marcy}, {Wright},
  {Driscoll}, {Butler}, {Hekker}, {Reffert}, \& {Vogt}}]{Johnson07}
---. 2007, \apj, 665, 785

\bibitem[{{Johnson} {et~al.}(2010{\natexlab{b}}){Johnson}, {Bowler}, {Howard},
  {Henry}, {Marcy}, {Isaacson}, {Brewer}, {Fischer}, {Morton}, \&
  {Crepp}}]{Johnson10c}
---. 2010{\natexlab{b}}, \apjl, 721, L153

\bibitem[{{Johnson} {et~al.}(2010{\natexlab{c}}){Johnson}, {Howard}, {Marcy},
  {Bowler}, {Henry}, {Fischer}, {Apps}, {Isaacson}, \& {Wright}}]{Johnson10}
---. 2010{\natexlab{c}}, \pasp, 122, 149

\bibitem[{{Malmquist}(1920)}]{Malmquist}
{Malmquist}, K. 1920, Medd.\ Lund.\ Astron.\ Obs., 22, 1

\bibitem[{{Marcy} {et~al.}(2005{\natexlab{a}}){Marcy}, {Butler}, {Fischer},
  {Vogt}, {Wright}, {Tinney}, \& {Jones}}]{Marcy_Japan_05}
{Marcy}, G., {Butler}, R.~P., {Fischer}, D., {Vogt}, S., {Wright}, J.~T.,
  {Tinney}, C.~G., \& {Jones}, H.~R.~A. 2005{\natexlab{a}}, Progress of
  Theoretical Physics Supplement, 158, 24

\bibitem[{{Marcy} {et~al.}(2000){Marcy}, {Butler}, \& {Vogt}}]{Marcy_saturns}
{Marcy}, G.~W., {Butler}, R.~P., \& {Vogt}, S.~S. 2000, \apjl, 536, L43

\bibitem[{{Marcy} {et~al.}(2005{\natexlab{b}}){Marcy}, {Butler}, {Vogt},
  {Fischer}, {Henry}, {Laughlin}, {Wright}, \& {Johnson}}]{Marcy05}
{Marcy}, G.~W., {Butler}, R.~P., {Vogt}, S.~S., {Fischer}, D.~A., {Henry},
  G.~W., {Laughlin}, G., {Wright}, J.~T., \& {Johnson}, J.~A.
  2005{\natexlab{b}}, \apj, 619, 570

\bibitem[{{Mayor} \& {Queloz}(1995)}]{Mayor_queloz}
{Mayor}, M., \& {Queloz}, D. 1995, \nat, 378, 355

\bibitem[{{Mayor} {et~al.}(2011){Mayor}, {Marmier}, {Lovis}, {Udry},
  {S{\'e}gransan}, {Pepe}, {Benz}, {Bertaux}, {Bouchy}, {Dumusque}, {Lo Curto},
  {Mordasini}, {Queloz}, \& {Santos}}]{Mayor11}
{Mayor}, M., {et~al.} 2011, arXiv:1109.2497

\bibitem[{{Pepe} {et~al.}(2002){Pepe}, {Mayor}, {Galland}, {Naef}, {Queloz},
  {Santos}, {Udry}, \& {Burnet}}]{Pepe02}
{Pepe}, F., {Mayor}, M., {Galland}, F., {Naef}, D., {Queloz}, D., {Santos},
  N.~C., {Udry}, S., \& {Burnet}, M. 2002, \aap, 388, 632

\bibitem[{{Perryman} \& {ESA}(1997)}]{PerrymanESA}
{Perryman}, M.~A.~C., \& {ESA}. 1997, {The HIPPARCOS and TYCHO catalogues.
  Astrometric and photometric star catalogues derived from the ESA HIPPARCOS
  Space Astrometry Mission} (The Hipparcos and Tycho catalogues.~Astrometric
  and photometric star catalogues derived from the ESA Hipparcos Space
  Astrometry Mission, Publisher: Noordwijk, Netherlands: ESA Publications
  Division, 1997, Series: ESA SP Series vol no: 1200, ISBN: 9290923997 (set))

\bibitem[{{Santos} {et~al.}(2001){Santos}, {Israelian}, \& {Mayor}}]{Santos01b}
{Santos}, N.~C., {Israelian}, G., \& {Mayor}, M. 2001, \aap, 373, 1019

\bibitem[{{Sato} {et~al.}(2005){Sato}, {Fischer}, {Henry}, {Laughlin},
  {Butler}, {Marcy}, {Vogt}, {Bodenheimer}, {Ida}, {Toyota}, {Wolf}, {Valenti},
  {Boyd}, {Johnson}, {Wright}, {Ammons}, {Robinson}, {Strader}, {McCarthy},
  {Tah}, \& {Minniti}}]{Sato05}
{Sato}, B., {et~al.} 2005, \apj, 633, 465

\bibitem[{{Tinney} {et~al.}(2001){Tinney}, {Butler}, {Marcy}, {Jones}, {Penny},
  {Vogt}, {Apps}, \& {Henry}}]{Tinney01}
{Tinney}, C.~G., {Butler}, R.~P., {Marcy}, G.~W., {Jones}, H.~R.~A., {Penny},
  A.~J., {Vogt}, S.~S., {Apps}, K., \& {Henry}, G.~W. 2001, \apj, 551, 507

\bibitem[{{Udry} {et~al.}(2003){Udry}, {Mayor}, {Clausen}, {Freyhammer},
  {Helt}, {Lovis}, {Naef}, {Olsen}, {Pepe}, {Queloz}, \& {Santos}}]{Udry03}
{Udry}, S., {et~al.} 2003, \aap, 407, 679

\bibitem[{{Vogt} {et~al.}(2002){Vogt}, {Butler}, {Marcy}, {Fischer},
  {Pourbaix}, {Apps}, \& {Laughlin}}]{Vogt02}
{Vogt}, S.~S., {Butler}, R.~P., {Marcy}, G.~W., {Fischer}, D.~A., {Pourbaix},
  D., {Apps}, K., \& {Laughlin}, G. 2002, \apj, 568, 352

\bibitem[{{Wright}(2005)}]{WrightErratum05}
{Wright}, J.~T. 2005, \aj, 129, 1776

\bibitem[{{Wright} {et~al.}(2009){Wright}, {Fischer}, {Ford}, {Veras}, {Wang},
  {Henry}, {Marcy}, {Howard}, \& {Johnson}}]{Wright09c}
{Wright}, J.~T., {et~al.} 2009, \apjl, 699, L97

\bibitem[{{Wright} {et~al.}(2011){Wright}, {Fakhouri}, {Marcy}, {Han}, {Feng},
  {Johnson}, {Howard}, {Fischer}, {Valenti}, {Anderson}, \& {Piskunov}}]{EOD}
---. 2011, \pasp, 123, 412

\end{thebibliography}

\end{document}